# Creating a Linked Data-Friendly Metadata Application Profile for Archival Description
*Poster*

Matienzo, Mark A.
Stanford University, U.S.A.
matienzo@stanford.edu

Roke, Elizabeth Russey
Emory University, U.S.A.
elizabeth.roke@emory.edu

Carlson, Scott
Rice University, U.S.A.
sjc5@rice.edu

## Abstract

We provide an overview of efforts to apply and extend Schema.org for archives and archival description. The authors see the application of Schema.org and extensions as a low barrier means to publish easily consumable linked data about archival resources, institutions that hold them, and contextual entities such as people and organizations responsible for their creation.

## Rationale and Objectives

Schema.org has become one of the most widely recognized and adopted mechanisms for publishing structured data on the World Wide Web, and has incorporated extensions to address the needs of specialist communities (Guha, et al., 2016). It has been used with some success in cultural heritage sector through libraries and digital collections platforms using both Schema.org core types and properties, as well as SchemaBibExtend, an extension for bibliographic information (bib.schema.org, n.d.). These uses include leveraging it as a means to improve search engine rankings (Scott, 2014), to publish library staff directories (Clark and Young, 2017) and to expose linked data about collections materials (Lampron, et al., 2016). However, the adoption of Schema.org in the context of archives has been somewhat limited.

Our project focuses on identifying pragmatic methods to publish linked data about archives, archival resources, and their relationships, and to identify gaps between existing models. In our initial round of work, we are looking at applying Schema.org as the core model, and are investigating and contributing to the proposed Schema Architypes extensions (W3C Schema Architypes Community Group, 2017e). We see this as an opportunity to demonstrate the potential of Schema.org as a minimally viable mechanism for publishing linked data about archives, their collections, and the entities involved in their creation and management. In addition, this project operates in the context of a larger area of effort, focused on providing archivists, metadata professionals, and technologists hands on experience with data model and ontology development.

We have identified a small number of key objectives for this initial round of work, including developing mappings to Schema.org and associated extensions such as Schema Architypes from archival description standards for search engine optimization and general web discovery; ideally producing RDF-modeled representations of archival description directly from archives management systems, rather than from representations exported from such systems (cf. Gracy, 2015); and alignment with other related data models and application profiles.

## Work Plan

Our work plan for the initial areas of investigation contains two phases. The first phase of the project, completed in April 2017, involved a survey of the landscape of related initiatives, and the identification of use cases which informed the objectives listed above. Our landscape survey focused on providing an initial review of potential models to serve as the basis for this work, including Schema.org; the Linking Lives project (Stevenson, 2012); the Bibframe Lite archives extension (Zepheira, n.d.; Zepheira and Atlas Systems, 2016); and the Europeana Data Model (Hennicke, et al., 2011). The group chose not to evaluate the draft Records in Context Conceptual Model (International Council on Archives Expert Group on Archival Description, 2016) for this purpose given its complexity, the lack of an associated ontology (originally scheduled to be released in late 2016), and the likelihood of substantial revision. Through this discussion, we decided to continue work on investigating a Schema.org profile for archives for four reasons: its simplicity and suitability towards both providing a basic representation of entities identified in archival description; the preexisting work on the Schema Architypes extension; the need to contribute domain expertise to the W3C Schema Architypes Community Group; and the opportunity to incorporate Schema.org markup directly into archival management and discovery applications, providing a representation suitable for search engines and easier consumption by other downstream client applications, such as request management systems. The Community group developed several modeling approaches (W3C Schema Architypes Community Group 2017a, 2017b, 2017c, 2017d; see figures 1 and 2) and submitted formal proposals to the Schema.org community in September 2017 (Wallis 2017a, 2017b; see figure 2). The types and properties introduced in the proposals are strong contenders to address our use cases related to search engine optimization, improved discovery, and consumption by client applications.

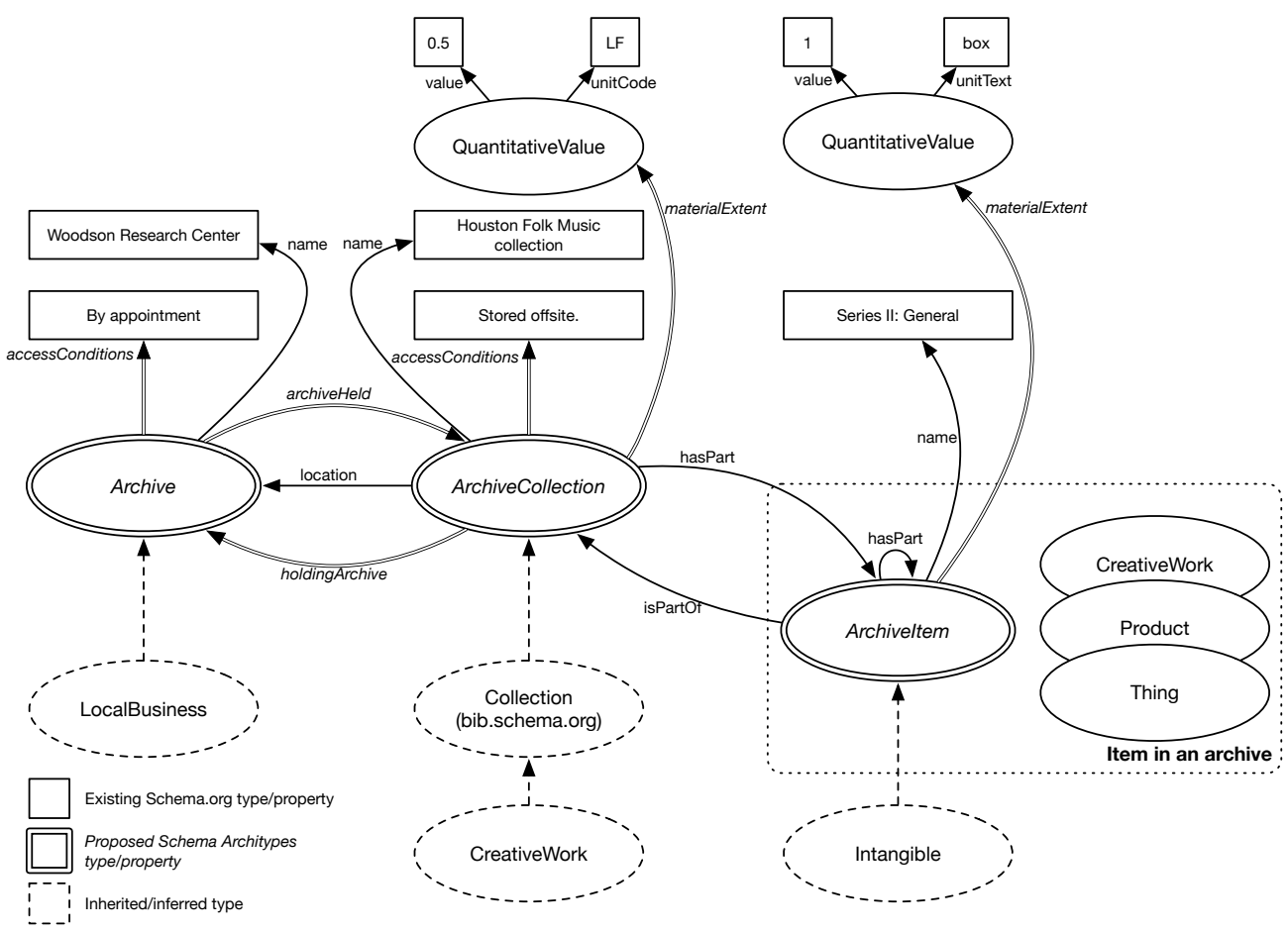

FIG. 1. Initial Schema Architypes proposal extension with extent extension. Adapted from W3C Schema Architypes Community Group 2017b, 2017d.

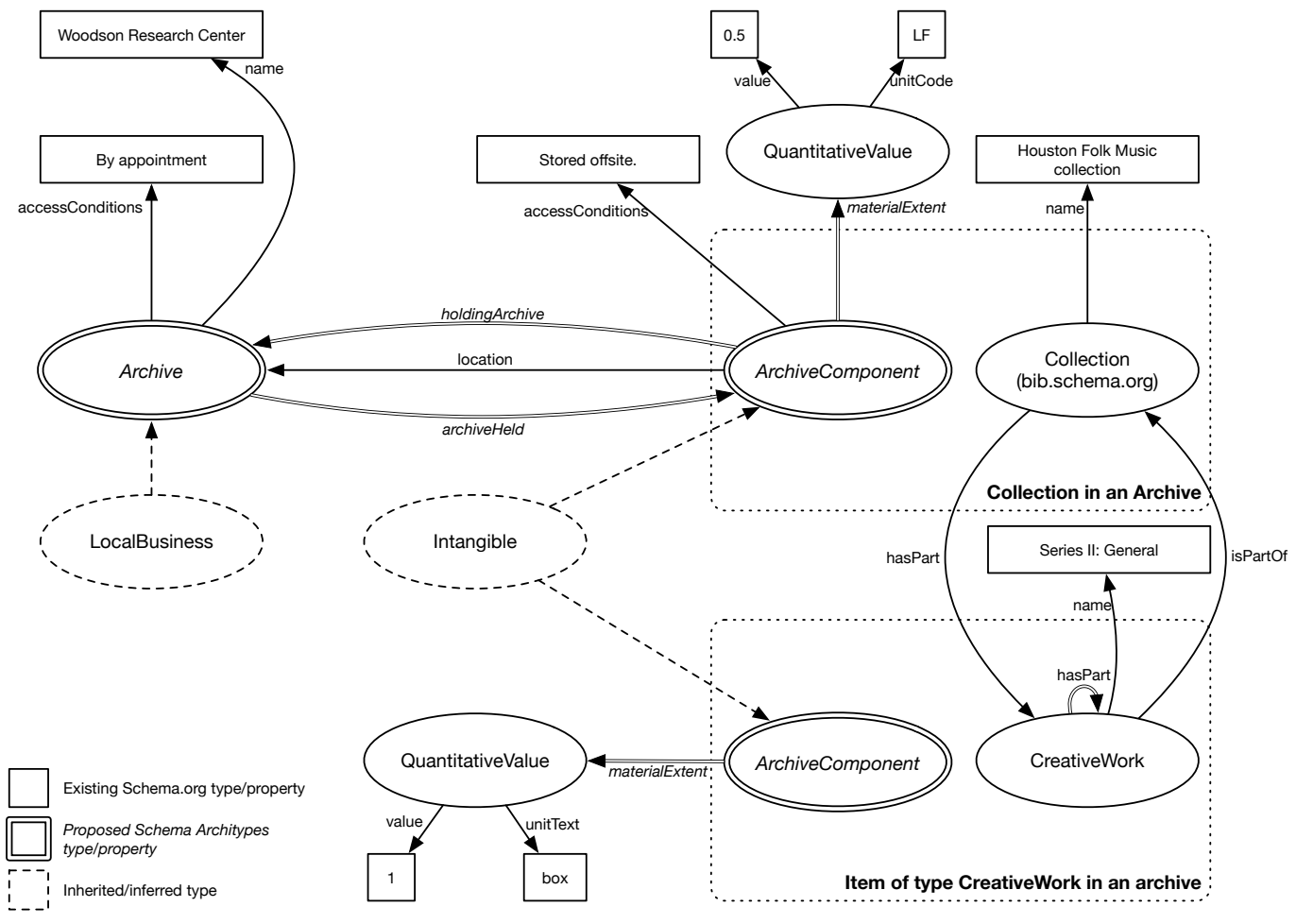

FIG. 2. Alternative Schema Architypes proposal with extent extension, submitted as Schema.org proposals (Wallis 2017a, 2017b). Adapted from W3C Schema Architypes Community Group 2017a, 2017b.

The second phase of the project, completed and pending feedback as of August 2017, was to undertake in-depth analysis of Schema.org and its associated extensions as a means to develop a profile suitable for publishing linked data for archives. Specific deliverables for this phase include identification of archival descriptive elements that should be expressed in Schema.org and undertaking a gap analysis of existing Schema.org and Schema Architypes types and properties; creating examples of Schema.org-based archival description; direct feedback and proposed revisions to the Schema Architypes Community Group; developing mappings from both content and structure standards for archival description (including ISAD(G), ISAAR-CPF, DACS, and Encoded Archival Description); and developing recommended mappings from data models of open source archives management applications such as ArchivesSpace (Matienzo and Kott, 2013) and AtoM (Artefactual Systems, 2015) to this profile. As of late May 2017, we have a completed a preliminary set of mappings from ISAD(G), ISAAR-CPF, DACS, and the ArchivesSpace and AtoM data models to Schema.org and the Architypes extensions for collection-level descriptions and information about agents and archival repositories, and have created a small number of draft examples used to verify our mappings. (See figure 3 for an example of DACS elements mapped to Schema extensions.)

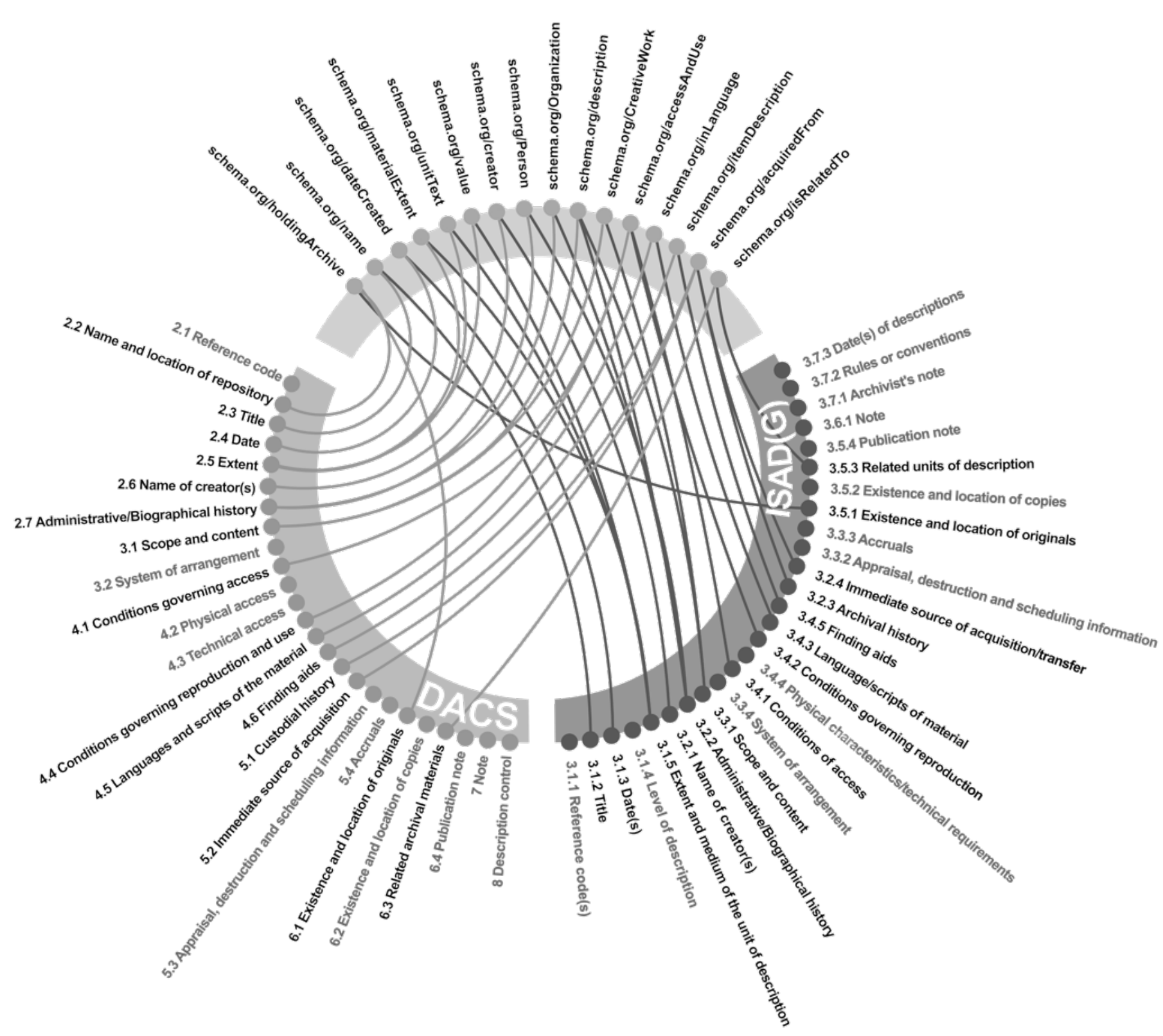

FIG. 3. Elements from *Describing Archives: A Content Standard* (bottom left) and *ISAD(G)* (bottom right) mapped to appropriate Schema.org properties. Note the *DACS* or *ISAD(G)* elements in gray, which do not yet have applicable Schema mapping.

The process of mapping existing descriptive standards and application-specific data models to Schema.org and the Schema Architypes extensions was mostly straightforward, with a few notable exceptions. The most substantive discussion within our group occurred around the desire or utility of mapping information in the description control area (e.g. *ISAD(G)* §3.7), which relates to information about the archival description itself, such as standards used, date of descriptions, and the like. After careful consideration, we chose to not to map this information and instead emphasized representation of collection metadata over metadata about finding aids. Our group also discussed the complications of mapping the level of description for a given unit (*ISAD(G)* §3.1.4), given a lack of consistency in existing practice and data, and a discussion about its direct relevance to addressing use cases around search engine optimization and improved discovery. These conversations led to a decision to not map this data despite its perceived importance by archivists. We believe this concern may be alleviated by using *isPartOf* and *hasPart* relationship properties expressed in Schema.org to emphasize contextual relationships across levels of description within an archival collection. Reference codes (*ISAD(G)* §3.3.1) were also identified as an area for further consideration given a lack of clarity in existing archival practice. While we investigated patterns developed for SchemaBibExtend for call numbers and barcodes (W3C Schema Bib Extend Community Group, 2015), we found these patterns to be ambiguous given widely varying practices in how reference codes are assigned or used by archival repositories. Beyond these areas, suitable mappings still have yet to be identified for information usually expressed as textual notes, such as information about appraisal, accruals, or arrangement (*ISAD(G)* §3.3.2-3.3.4); physical characteristics and technical requirements related to access (*ISAD(G)* §3.4.4); and references to originals or copies of archival material (*ISAD(G)* §3.5.1-3.5.2). We expect additional feedback from archivists, metadata professionals, and other stakeholders will allow us to identify candidate mappings for these gaps, and will provide the necessary feedback to validate or refine our analysis.

## Expected Benefits and Future Work

Our project provides a satisfactory proof of concept and test corpus of information about archives that will serve as a basis for fuller implementations. We believe that this will additionally allow institutions to better understand limitations in their existing descriptive data. To that end, the group is actively soliciting additional examples of archival description expressed using Schema.org and the proposed extensions (Archives and Linked Data Interest Group, 2017). Given our focus in mapping from archives management systems to a profile based on Schema.org and Schema Architypes, we see the opportunity to implement this profile directly in applications designed to support discovery of archival information, such as the public user interfaces provided by management systems like ArchivesSpace and AtoM, as well as other open source archival discovery-focused applications such as staticAid (Arnold, et al., 2017) and ArcLight (Stanford University Libraries, 2017). In addition, we expect to extend our work to undertake more in-depth investigation of and mapping to other proposed ontologies and data models for archives, with the possibility of generating extension ontologies or application profiles through further gap analysis.

## Acknowledgements

Special thanks to the members of the Archives and Linked Data interest group working on this project: Scott Carlson, Mark Custer, Patrick Galligan, Dan Gillean, Gloria Gonzalez, Maggie Hughes, Mark Matienzo, Dave Mayo, Laney McGlohon, Evelyn McLellan, Katy Rawdon, and Elizabeth Russey Roke.